\documentclass[%
  amsmath,amssymb,
  aip, jap, preprint,
  ]{revtex4-1}

  \usepackage{graphicx}
  \usepackage{dcolumn}
  \usepackage{bm}
  \usepackage[normalem]{ulem} 
  \usepackage{acro}
  \usepackage{color}

\DeclareAcronym{BZ}{
  short=BZ,
  long=Brillouin zone,
  }
\DeclareAcronym{TWS}{
  short=TWS,
  long=topological Weyl semimetal
  }
\DeclareAcronym{TDS}{
  short=TDS,
  long=topological Dirac semimetal
  }
\DeclareAcronym{3D}{
  short=3D,
  long=three-dimensional,
  }
\DeclareAcronym{2D}{
  short=2D,
  long=two-dimensional,
  }
\DeclareAcronym{SOI}{
  short=SOI,
  long=spin-orbit interaction,
  }
\DeclareAcronym{ARPES}{
  short=ARPES,
  long=angular resolved photoemission spectroscopy,
  }
\DeclareAcronym{DFT}{
  short=DFT,
  long=density functional theory,
  }
\DeclareAcronym{XC}{
  short=XC,
  long=exchange and correlation,
  }
\DeclareAcronym{PBE}{
  short=PBE,
  long={Perdew, Burke and Ernzerhof},
  }
\DeclareAcronym{GGA}{
  short=GGA,
  long=generalized gradient approximation,
  }
\DeclareAcronym{PW}{
  short=PW,
  long=plane wave,
  }
\DeclareAcronym{EELS}{
  short=EELS,
  long=electron energy loss spectra,
  }
\DeclareAcronym{bct}{
  short=bct,
  long=body-centered tetragonal,
  }
\DeclareAcronym{QE}{
  short=QE,
  long=QUANTUM ESPRESSO,
  }
\DeclareAcronym{DOS}{
  short=DOS,
  long=density of states,
  }
\DeclareAcronym{IR}{
  short=IR,
  long=infrared,
  }
\DeclareAcronym{UV}{
  short=UV,
  long=ultraviolet,
  }

\begin{document}
\title{
  \emph{Ab initio} optical and energy loss spectra of transition metal monopnictides TaAs, TaP, NbAs and NbP
  }

\author{
  Davide Grassano$^{1}$,
  Friedhelm Bechstedt$^{2}$,
  and Olivia Pulci$^{1}$
  }
\affiliation{
  $^1$ Dipartimento di Fisica, Universit\`{a} di Roma Tor Vergata, and INFN, Via della Ricerca Scientifica 1, I-00133 Rome, Italy\\
  $^2$ Institut f\"{u}r Festk\"{o}rpertheorie und -optik, Friedrich-Schiller-Universit\"{a}t, Max-Wien-Platz 1, 07743 Jena, Germany
  }

\date{\today}

\begin{abstract}
  Transition metal monopnictides represent a new class of topological semimetals with low-energy excitations, namely, Weyl fermions.
  We report optical properties across a wide spectral energy range for TaAs, TaP, NbAs and NbP, calculated within density functional theory.
  Spectra are found to be somewhat independent of the anion and the light polarization.
  Their features are explained in terms of the upper  $s$, $p$, $d$, and $f$ electrons.
  Characteristic absorption features are related to the frequency dependence of the Fresnel reflectivity.
  While the lower part of the energy loss spectra is dominated by plasmonic features, the high-energy structures are explained by interband transitions.
\end{abstract}

\maketitle

\section{Introduction}\label{sec:intro}
  Transition metal monopnictides belong to a class of materials known as \Acp{TWS}, which have recently attracted great interest \cite{young.zaheer.2012,armitage.mele.2018}. 
  One characteristic property of  a Weyl semimetal is the presence of low-energy electronic excitations that can be described by a \ac{3D} Weyl Hamiltonian.
  Band crossings, known as Weyl nodes, occur near the Fermi energy. 
  \Acp{TWS} can be considered as  \aclp{TDS} \cite{wang.fang.2012,conte.pulci.2017}, where either the time-reversal or inversion symmetry has been broken. 
  The symmetry lowering causes a splitting of one Dirac point into two Weyl nodes \cite{burkov.balents.2011,burkov.hook.2011} either in energy or momentum space. 
  The inclusion of \ac{SOI} in the calculations is fundamental to achieve the correct crossings and splittings.

  The presence of the Weyl nodes confers these materials with novel properties such as Fermi arcs of surface states \cite{wang.fang.2012,wan.turner.2011}, anomalous and spin Hall conductivity \cite{chang.zhang.2013,yu.zhang.2010,murakami.nagosa.2003,sinova.culcer.2004,rauch.mertig.2017}, negative magnetoresistance \cite{huang.zhao.2015}, and the Adler-Bell-Jackiw chiral anomaly \cite{nielsen.nunomiya.1983}.
  Interesting applications of \acp{TWS} as materials for Veselago lenses have been predicted \cite{hills.kusmartseva.2017}.

  In order to realize a \ac{TWS}, several approaches have been proposed. 
  One way is to break time-reversal symmetry by magnetic doping of Dirac semimetals and magnetic ordering \cite{burkov.balents.2011}.
  Another possibility  is to break the inversion symmetry by appropriate atomic arrangements.
  Transition metal monopnictides crystallize in the \ac{bct} structure with the non-symmorphic space group I4$_1$md, which naturally lacks inversion symmetry \cite{huang.xu.2015}.
  Single crystals of the TaAs class can be grown through chemical vapor transport techniques using iodine as the transport medium.
  Optimization of the growth procedure leads to single crystals larger than $1$ cm in size\cite{li.2016}.

  Experimental realization of  \acp{TWS} spurred  interest into a deeper analysis of their properties.
  The low-energy excitations of these materials, in particular their Weyl nature, have been amply studied theoretically \cite{huang.xu.2015,weng.fang.2015,sun.wu.2015,lee.xu.2015,grassano.pulci.2018}
  and experimentally \cite{xu.belopolski.2015.TaAs,lv.weng.2015,xu.belopolski.2015.TaP,xu.alidoust.2015,souma.wang.2016,xu.dai.2016,kimura2017optical}. 
  However, there is still limited knowledge of the visible and \ac{UV} optical responses of Weyl materials, which is needed to characterize their crystal quality and electronic structure, as well as adapt them for use in  opto-electronic devices.
  Therefore, a better understanding of their optical and energy-loss spectra in a wide energy range is critical for the practical application of transition metal monopnictides.

  Our paper is organized as follows:
  In Sec. \ref{sec:methods} we describe the theoretical methods employed.
  In Sec. \ref{sec:results} we present and discuss the calculated optical and energy loss spectra.
  In Sec. \ref{sec:conclusions} we give a summary and conclusions.

\section{Theoretical methods}\label{sec:methods}
  The structural, electronic and optical properties of the transition metal monopnictides have been computed using \ac{DFT} as implemented in the \ac{QE} Package \cite{espresso.2009,espresso.2017}.
  The \ac{XC} potential was generated within the \ac{GGA} scheme, as formulated by \ac{PBE} \cite{PBE.1996}.
  The following electronic configuration have been employed in order to generate fully relativistic, norm-conserving pseudopotentials\cite{hamann.2013}: Ta(4$f^{14}$5$s^2$5$p^6$5$d^3$6$s^2$), Nb(4$s^2$4$p^6$4$d^3$5$s^2$), As(3$d^{10}$4$s^2$4$p^3$) and P(3$s^2$3$p^3$). 
  The $4f$ states of Ta and $3d$ states of As have been included since they fall in the analyzed energy range.
  The cut-off radii and reference channels have been chosen so that the pseudo-wavefunctions agree  with the all-electron ones, without the presence of any ghost states \cite{gonze.kackell.1990}.
  The \ac{SOI} has been included by solving the radial Dirac equation for each isolated atom and then, by reducing the four-component Dirac spinors to two-component Pauli spinors \cite{corso.conte.2005,conte.fabris.2008}.
  Optical and energy loss spectra have been computed using the Yambo code \cite{yambo.2009} taking into account the non-locality of the pseudopotentials\cite{delsole.girlanda.1993}.

  Atomic geometries have been optimized starting from the reference Wyckoff positions for the I4$_1$md(No. 109) group with multiplicity 4: $(0,0,u)$ and $(0,1/2,u+1/4)$ with $u=0$ for the cation and $u\neq0$ for the anion \cite{wondratsheck.2004}.
  Convergence with the number of \acp{PW} has been tested, yielding  a suitable energy cut-off of $100$ Ry.
  The \ac{BZ} of the \ac{bct} has been sampled  using a uniform grid of $21 \times 21 \times 7$ Monkhorst-Pack ${\bf k}$ points \cite{monkhorst.pack.1976} in the self-consistent calculations. 
  The energy eigenvalues and eigenfunctions of the single-particle Kohn-Sham equation are used to characterize the electronic properties of the \acp{TWS}.

  Convergence for the optical properties in the energy range from $0.2$ to $50 \ eV$ required instead a $36 \times 36 \times 12$ ${\bf k}$ point grid, with $190$ empty bands.
  For the low-energy range a much more dense  grid of $192 \times 192 \times 64$  ${\bf k}$ points cropped around the Weyl nodes  has been used.
  The optical properties have been computed within the single-particle approximation in the transverse gauge.
  The imaginary part of a diagonal element $\epsilon _{jj} \left( \omega \right)$ of the frequency-dependent dielectric tensor is given as

  \begin{gather}\label{eq:opt_transvers}
  \begin{split}
    {\rm Im}\epsilon_{jj} \left( \omega \right)
      = & \left(
            \frac{2\pi e}
                 {m\omega}
          \right)^2
          \frac{1}
               {V}
          \sum_{\bf k}
            \sum_{c,v}
              \left[
                f\left(\varepsilon_v({\bf k})\right) - 
                f\left(\varepsilon_c({\bf k})\right)
              \right]
    \times \\
        & \times \left|
            \langle 
              c{\bf k}|p_j|v{\bf k}
            \rangle
          \right|^2
          \delta
          \left(
            \varepsilon_c({\bf k}) -
            \varepsilon_v({\bf k}) -
            \hbar\omega
          \right) ,
  \end{split}
  \end{gather}
  where  Bloch states $\left| n {\bf k} \right>$  with band index $n = c,v$, wavevector ${\bf k}$, energy $\varepsilon _{n} \left( {\bf k} \right)$ and occupation $f \left( \varepsilon _{n} \left( {\bf k} \right) \right)$ were used to compute the transition matrix elements of the momentum operator $p_j$ in $j$-th Cartesian direction between valence ($v$) and conduction ($c$) bands.
  The mass symbol $m$ and the elementary charge $e$ are used in \eqref{eq:opt_transvers}.
  The effects of the non-local part of the pseudopotentials \cite{delsole.girlanda.1993} on the momentum matrx elements have been included by computing the dipole corrections \cite{starace.1971} as implemented in the Yambo package\cite{yambo.2009}.
  The real part of the dielectric function has been obtained by applying the Kramers-Kronig relation \cite{book.lucarini.2005}

  \begin{gather}\label{eq:kk}
  \begin{split}
    {\rm Re}~\epsilon _{jj}(\omega) 
      &= 1 + \frac{2}
              {\pi}
         P\int_{0}^{\infty} 
         \frac{\omega ^\prime {\rm Im}(\epsilon _{jj}(\omega ^\prime))}
              {\omega ^{\prime 2}-\omega^2} d\omega ^\prime \
  \end{split}
  \end{gather}
  with $P$ as the principal value.

  The energy loss spectrum $L _j \left( \omega \right)$ is calculated for vanishing momentum transfer as: 
  \begin{gather}\label{eq:eels}
  \begin{split}
    L _j \left( \omega \right) 
      & =  -{\rm Im}~\frac{1}
                   {\epsilon _{jj} \left( \omega \right)} \ ,
  \end{split}
  \end{gather}
  where $\hbar \omega$ represents the loss energy.
  The plasma frequencies for the free electron approximation are calculated using the formula
  \begin{gather}\label{eq:free_plasmon}
  \begin{split}
    \omega _p = \sqrt[]{\frac{4 \pi n_e e^2}{m}}
  \end{split}
  \end{gather}
  with $n_e$ as the density of the electrons contributing to the loss.
  This expression will only be  applied to discuss high-energy spectra above $20$ eV

\section{Results and discussion}\label{sec:results}
  \subsection{Atomic and electronic structures}\label{sec:structures}
    The transition metal monopnictides here studied crystallize in a body centered tetragonal structure. 
    Each anion (As, P) coordinates with 6 cations (Ta, Nb), and vice versa. 
    The   local geometry of the bonds is trigonal prismatic, due to a $spd^4$ hybridization.

    The equilibrium atomic geometries have been obtained by relaxing the atomic positions while varying the lattice parameters $a$, $c$ and $u$. 
    The calculated structures\cite{grassano.pulci.2018} are in agreement (with deviations smaller than 1\%) with results from X-ray diffraction measurements\cite{furuseth.selte.1965,boller.parthe.1963,schonberg.1954} and other theoretical calculations \cite{lee.xu.2015}.

\begin{figure*}[!ht]
      \includegraphics[width=1.0\linewidth]{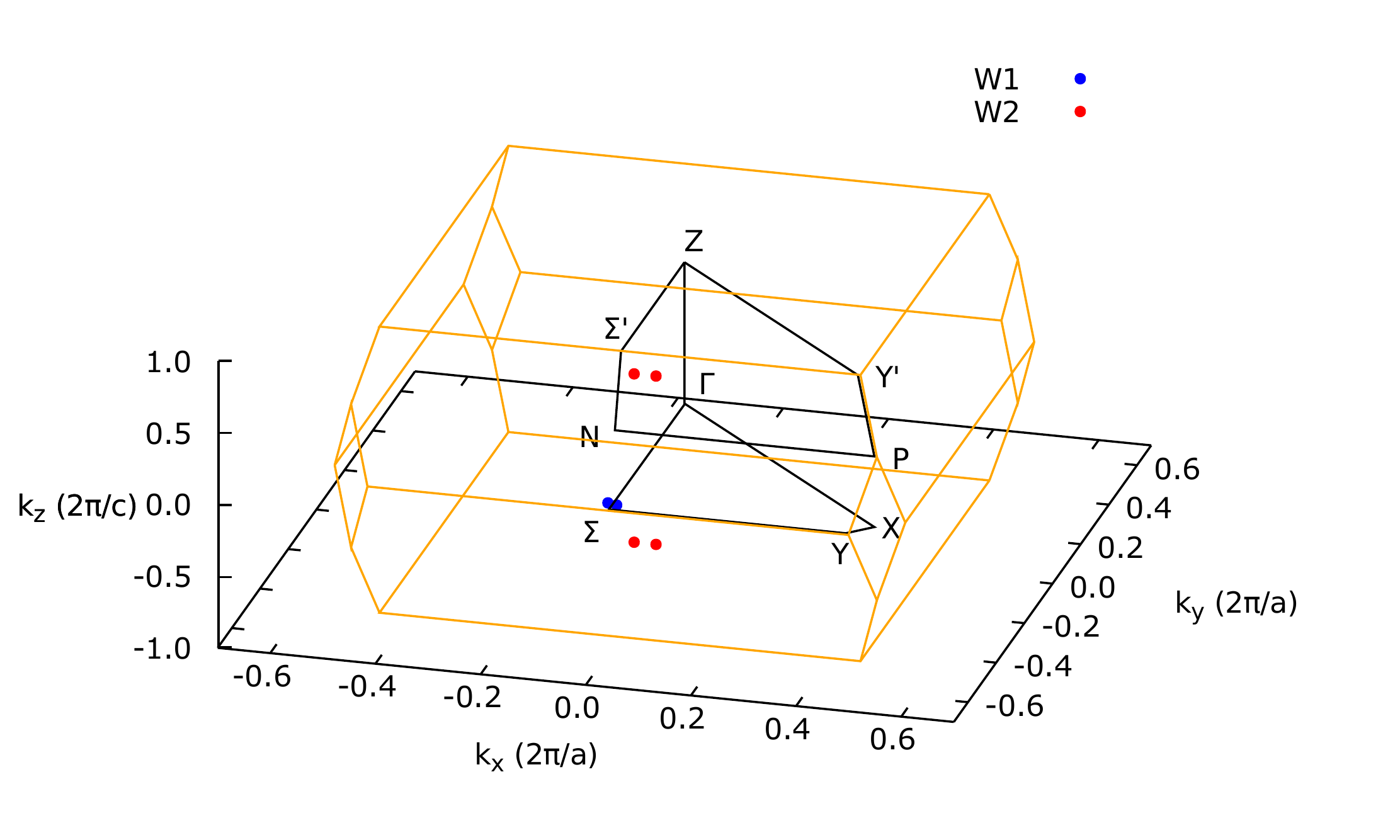}
      \caption{\label{fig:BZ} Brillouin zone (orange lines) of \ac{bct} TaAs. High-symmetry lines (black) and Weyl nodes, W1 (blue dots) and W2 (red dots) are shown.
      For clarity, only three pairs, one W1 and two W2, are shown in proximity of the $k_x = 0$ plane with the restriction $k_y < 0$.}
    \end{figure*}

\begin{figure*}[!ht]
      \includegraphics[width=1.0\linewidth]{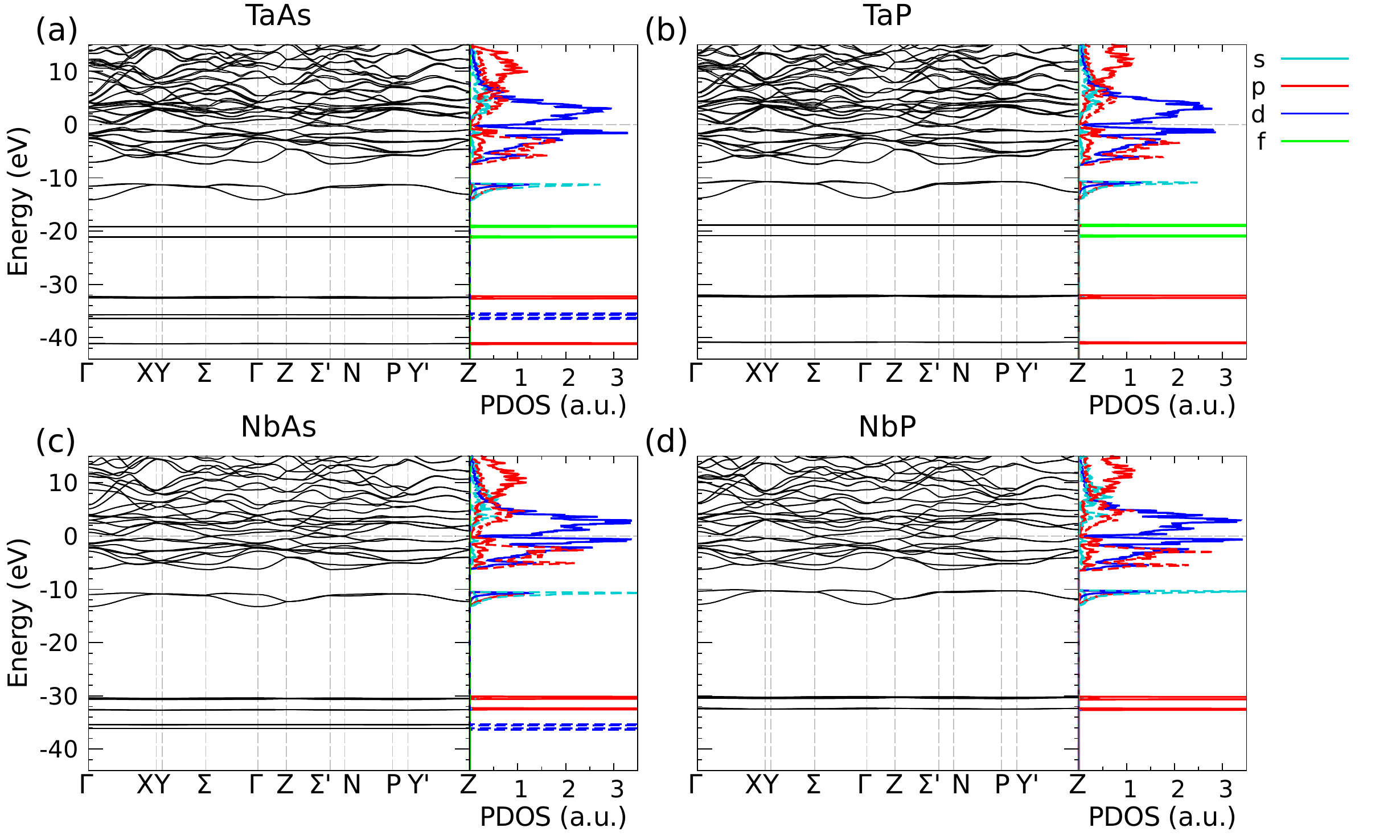}
      \caption{\label{fig:bands} Band structure along the high-symmetry directions for (a) TaAs, (b) TaP, (c) NbAs and (d) NbP. The projected \ac{DOS} (in arbitrary units) shows the contribution of the cation (solid lines) and of the anion (dashed lines). 
      The states $s$ (cyan), $p$ (red), $d$ (blue) and $f$ (green) are resolved.
      The Fermi level is used as energy zero.}
    \end{figure*}

    The band structures along the high-symmetry directions in the bct \ac{BZ}, highlighted in Fig. \ref{fig:BZ}, have been computed for all materials and are displayed in Fig. \ref{fig:bands}. 
    Weyl nodes are not visible since they lie away from the high-symmetry directions.
    Whereas the W1 nodes are located in on the $k_z = 0$ mirror plane close to the $\Gamma \Sigma$ direction, the W2 nodes occur far from any high-symmetry lines. The closest ones appear near the $\Gamma \Sigma ^\prime$ and $\Gamma N$ lines.
    It is evident from the projected \ac{DOS} data that the inclusion of the As $3d$ and Ta $4f$ electrons in the pseudopotential is required in order to obtain the correct results for the optical properties in a wide energy range.
    We see that the $d$ states of the cations give the dominant contributions to the bands close to the Fermi level. 
    In general, the displayed valence and conduction bands are built by cation $d$ and anion $p$ states with smaller contributions from the cation $s$ and $p$.
    The separated valence bands with binding energy slightly larger than $10$ eV are built by cation $d$ and $p$ states and anion $s$ states.
    Many deeper core-level bands are also visible in Fig. \ref{fig:bands}.
    They are due to Nb $4p$ and Ta $5p$ states below $-30$ eV, which are split by strong \ac{SOI}.
    Around $-20$ eV the \ac{SOI}-split Ta $4f$ bands appear in Fig. \ref{fig:bands}a and b.
    In Fig. \ref{fig:bands}a and c the As $3d$ derived bands are observable at about $-35$ eV.

  \subsection{Optical properties}\label{sec:eps}
\begin{figure*}[!ht]
      \includegraphics[width=1.0\linewidth]{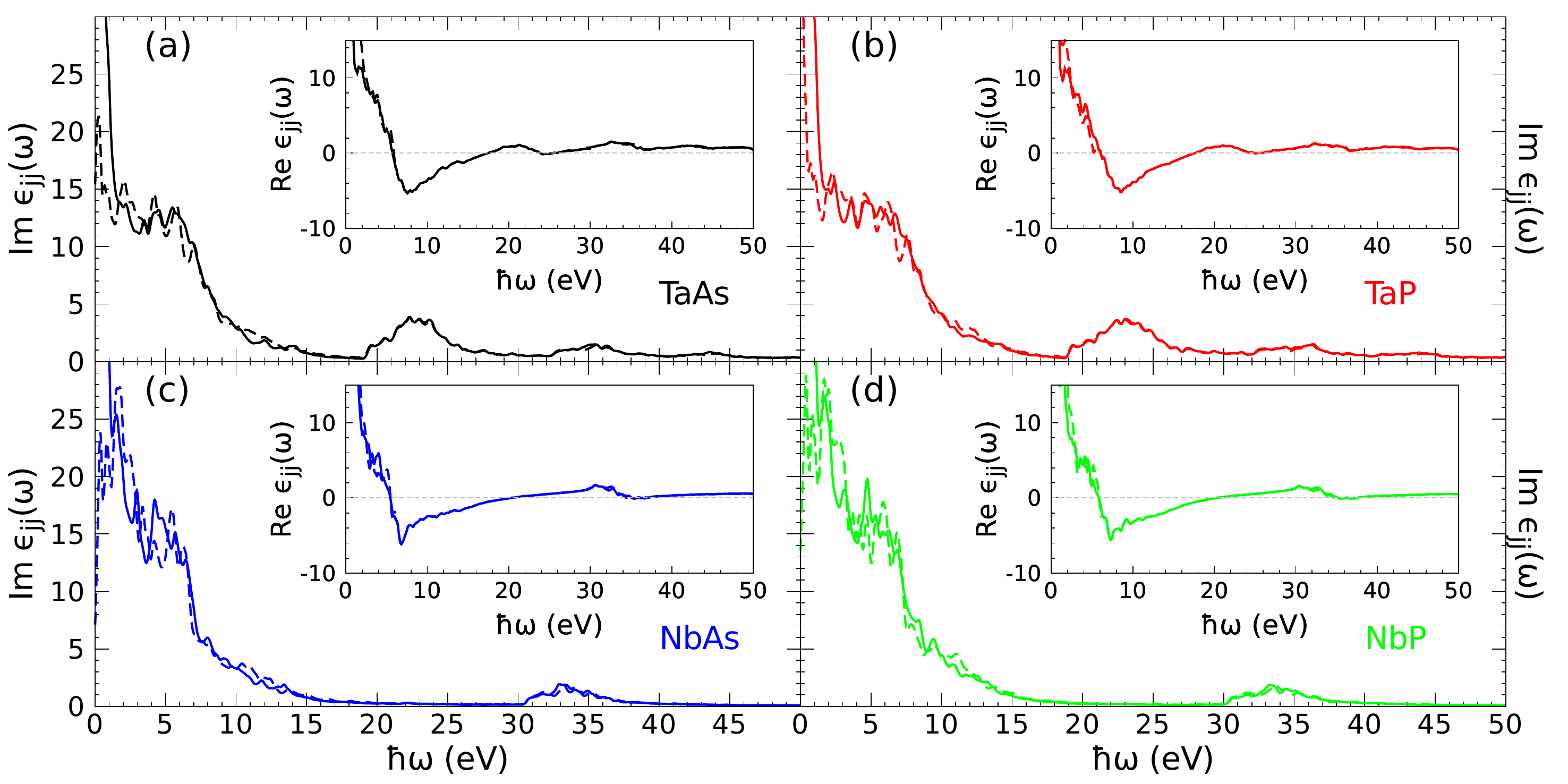}
      \caption{\label{fig:eps2} 
      Imaginary part of the dielectric function for $j=x,y$ polarization (solid lines)  and $j=z$ polarization (dashed lines) for TaAs (black), TaP (red), NbAs (blue) and NbP (green). 
      Inset: real part of the dielectric function. }
      \end{figure*}

    Using a dense mesh of {\bf k} points, we calculate the imaginary part of the dielectric function ${\rm Im}\ \epsilon_{jj} \left( \omega \right)$. 
    The results are reported in Fig. \ref{fig:eps2}, where convergence with the number of {\bf k} points and empty bands has been carefully tested.

    We observe that the absorption spectra ${\rm Im}\ \epsilon_{jj} \left( \omega \right)$ are fairly isotropic across a wide range of energies.
    Only in the low energy range (below $5$ eV) the tetragonal anisotropy of the crystal becomes evident as shown by the 
    dependence on the light polarization.
    The strong anisotropy in the infrared spectral range has been demonstrated in a previous paper \cite{grassano.pulci.2018}.

    It is worth noticing the similarity between the optical spectra for the same cation and, hence, the weak influence of the anion.
    The high-energy range is dominated by the cation as demonstrated by comparing   the TaAs/TaP (NbAs/NbP) spectra in Fig. \ref{fig:eps2}, while the anion  only gives rise to small intensity changes for energies lower than $15$ eV.
    The distinct features that characterize the high-energy transitions for TaX (X=As,P), visible in the range $19-27$ eV, are caused by transitions between the $f$ states of Ta and the conduction bands. 
    For all materials a broad peaked structure, visible in the range $30-38$ eV, is caused by transitions originating from the $p$ orbitals of the cation, visible around $-30$ eV in the band structure in Fig.~\ref{fig:bands}.
    In the case of TaX, another structure is visible at $43$ eV. 
    This  also originates from transitions from the $p$ orbitals, which, in the case of TaX, produce two structures instead of one because of the large \ac{SOI} split. 
    In contrast, the spectral feature at $43$ eV is not present in NbX (X=As,P) due to the smaller spin-orbit splitting.

    The reflectivity at normal incidence can also be calculated from the knowledge of the dielectric function via the Fresnel formula
    \begin{equation}
      R _j \left( \omega \right) =
        \left|
          \frac{\sqrt[]{\epsilon_{jj} \left( \omega \right)} - 1}
               {\sqrt[]{\epsilon_{jj} \left( \omega \right)} + 1}
        \right|
        ^2 \ .
    \end{equation}

    \begin{figure*}[!ht]
      \includegraphics[width=0.9\linewidth]{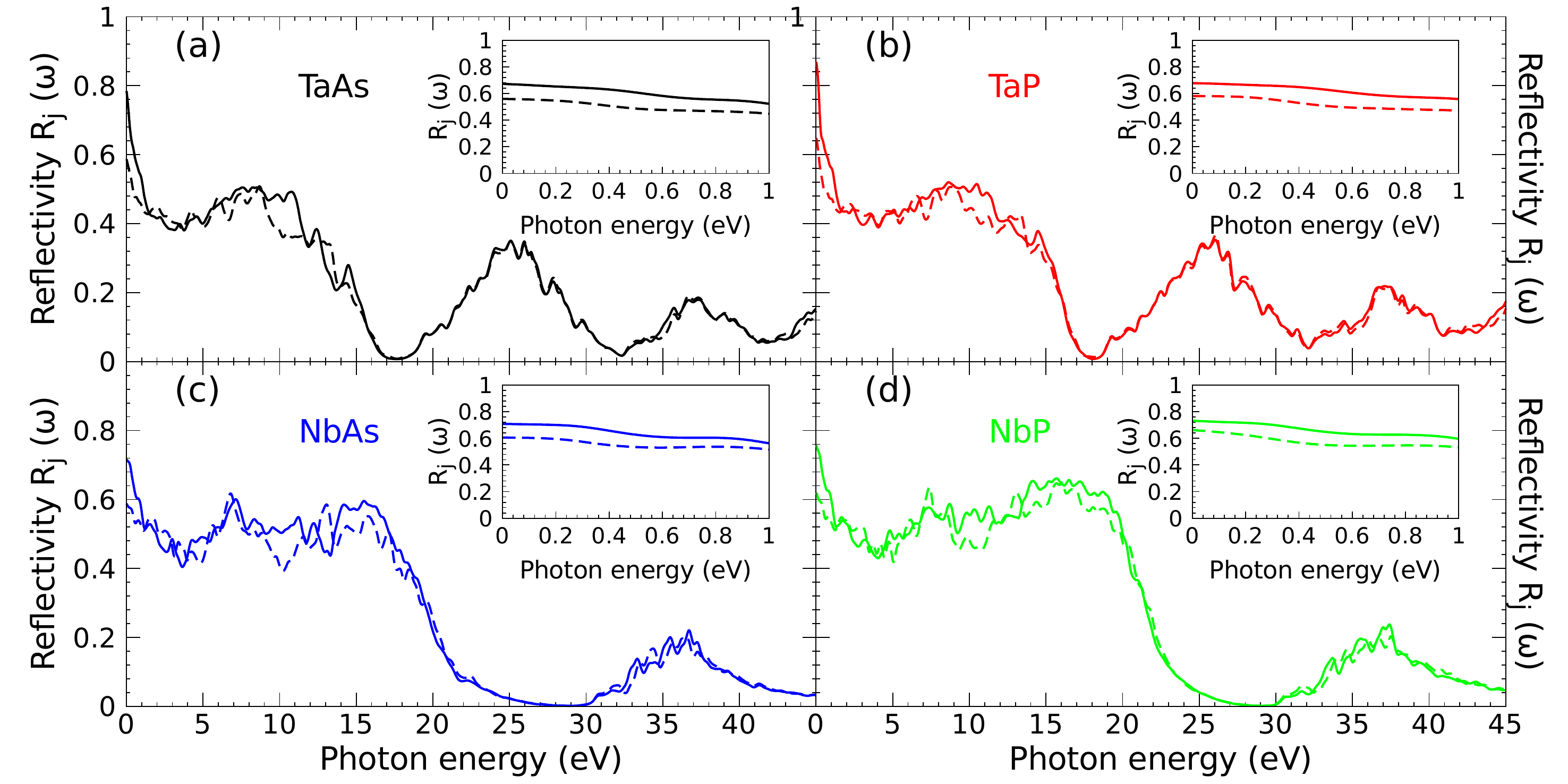}
      \caption{\label{fig:refl} Reflectivity for $j=x,y$ polarization (solid lines)  
      and for $j=z$ polarization (dashed lines) for TaAs (black), TaP (red), NbAs (blue) and NbP (green).
      Inset:  infrared reflectivity for  $x$, $y$ and $z$ polarizations.}
    \end{figure*}
    The reflectivity spectra presented in Fig. \ref{fig:refl} in the wide energy range show several structures that are connected to the same transitions as described for ${\rm Im}\ \epsilon_{jj} \left( \omega \right)$. 
    In particular we note that the reflectivity is negligible around $18$ eV ($29$ eV) for TaX (NbX) and X=As, P, where also ${\rm Im}\ \epsilon_{jj} \left( \omega \right)$ vanish.
    In these energy windows the transmission is hence maximized.
    In the intermediate energy range, $5-15$ eV for TaX and $5-18$ eV for NbX, the reflectivity is less frequency dependent, with values around $R_j \left( \omega \right) \approx 0.5-0.6$.
    Only below 5 eV down to  vanishing photon energy, a strong increase of the reflectivity is visible. 
    In this spectral range, the transition metal monopnictides reflect but also strongly absorb light.
    Finally, the infrared region (insets in Fig. \ref{fig:refl}),  characterizes the semimetallic character of the monopnictides with vanishing gaps and topological electron and hole pockets but also trivial hole pockets\cite{grassano.pulci.2018}.
    Caution should be taken when relating the low-energy behavior to the Weyl fermion picture. 
    In fact, although in the infrared spectral region an ideal Weyl node would give a constant contribution to the imaginary part of the dielectric function, the non-local dependence shown in equation \eqref{eq:kk} for the derivation of ${\rm Re}\ \epsilon$ does not allow to guess any special behavior in the infrared limit of the reflectivity.
    Caution is also necessary when comparing with experimental studies for photon energies below $0.1$ eV. 
    Real semimetals show influences of polar optical phonons and intraband electronic transitions \cite{kimura2017optical}.
    Consequently, the experimental finding that $R \rightarrow 1$ for $\hbar \omega \rightarrow 0$, due to the Drude contributions, cannot be reproduced in the insets of Fig. \ref{fig:refl}.

    Above the vanishing reflectivities, around $18$ and $29$ eV, respectively, further spectral features with intermediate reflectivity strength appear in Fig. \ref{fig:refl}.
    They are strongly related to the spectral features in the absorption spectra in Fig. \ref{fig:eps2}.
    They are consequences of the same interband transitions.
    The finite reflectivities in the high-energy range are important if the real parts of the optical conductivity, even in the infrared region, are constructed from measured reflectivity spectra and Kramers-Kronig relations\cite{xu.dai.2016}.

  \subsection{Energy losses}\label{sec:eels}
\begin{figure}[!ht]
      \includegraphics[width=0.9\linewidth]{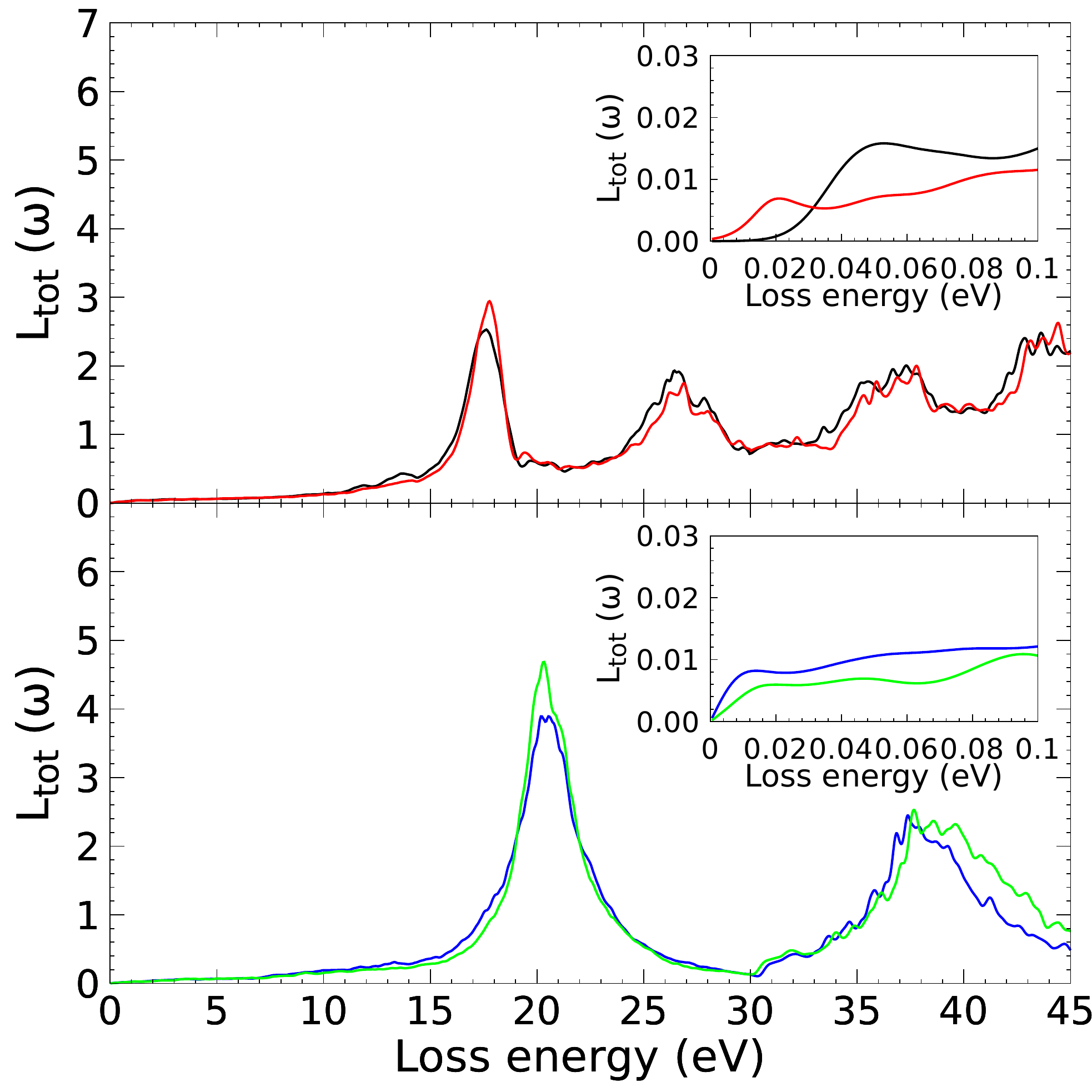}
      \caption{\label{fig:eels} Energy loss spectra for TaAs (black), TaP (red), NbAs (blue) and NbP (green). 
      Insets: low-energy range of the same quantities}
    \end{figure}

    By implementing Eq. \eqref{eq:eels} we calculate energy loss spectra for zero exchanged momentum.
    Only averaged values are shown in Fig. \ref{fig:eels} since the polarization dependence of the response functions is very weak.
    We also display low-energy spectra (Fig. \ref{fig:eels} insets) which provide useful information about the Weyl nodes, when the contributions of the trivial points are separated.
    In this case, low-electron energy loss spectroscopy is suggested as a complementary experimental method to study Weyl fermions.

    The first main peaks in the energy loss spectra are a consequence of the second zeros of ${\rm Re}\ \epsilon_{jj} \left( \omega \right)$ when ${\rm Im}\ \epsilon_{jj} \left( \omega \right)$ simultaneously nearly vanishes (see insets of Fig. \ref{fig:eps2}).
    The spectra show typical plasmon losses of valence electron gases as indicated by the positions near $\hbar \omega _p$, which are in the range of the plasmons of TaAs and NbAs compounds ($\sim 18$ eV) and TaP and NbP ($\sim 19$ eV), if the  $d^3s^2$ electrons of the cation and $p^3$ electrons of the anion are considered as free electrons.
    For all compounds these values do not vary with the propagation direction of the inelastically scattered particles.

    While the first main peak can be explained within the free electron plasmon picture \eqref{eq:free_plasmon}, the other high-energy structures that appear in Fig. \ref{fig:eels} are related to the optical transitions from semi-core states to the conduction bands as already discussed for ${\rm Im}\ \epsilon _{jj} \left( \omega \right)$.
    In particular for TaX we see transitions from the $f$ cation states associated with the 2nd peak at $28$ eV. 
    The 3rd and 4th peak (around $38$ and $44$ eV) are instead associated with transitions from the \ac{SOI}-split $p$ states of the cation.
    For NbX only another peak appears at $38$ eV, associated with transitions from the $p$ states of the cation.

\section{Summary and conclusions}\label{sec:conclusions}
  Results for the optical properties of transition metal monopnictides TaAs, TaP, NbAs and NbP, obtained $ab \ initio$ via \ac{DFT} and the independent particle approximation were presented across a wide range of photon energies. 
  The investigations of the projected \ac{DOS} and optical properties show the importance of the $f$ states of Ta, while the $d$ states of As give negligible contributions.
  The energy loss and reflectivity spectra were investigated as possible tools to further characterize the structural and electronic properties of  topological semimetals.
  The high-energy peaks in the optical properties are related to transitions from cation $f$ and $p$ semi-core states to conduction bands.
  All monopnictides show a high optical reflectivity in the range $0-15$ eV of $50-60\%$, while it still amounts to $20 \%$ around $35-40$ eV.
  The first peak of the energy loss is explainable through a plasmon picture by taking into account the cation $d^3 s^2$ and anion $p^3$ electrons as free particles.

\begin{acknowledgments}
  O.P. acknowledges financial funding from the EU project CoExAN
  (GA644076). F.B. acknowledges travel support by INFN Tor Vergata. CPU time was granted by CINECA HPC center.
\end{acknowledgments}

\bibliography{bibliography}

\begin{thebibliography}{44}%
\makeatletter
\providecommand \@ifxundefined [1]{%
 \@ifx{#1\undefined}
}%
\providecommand \@ifnum [1]{%
 \ifnum #1\expandafter \@firstoftwo
 \else \expandafter \@secondoftwo
 \fi
}%
\providecommand \@ifx [1]{%
 \ifx #1\expandafter \@firstoftwo
 \else \expandafter \@secondoftwo
 \fi
}%
\providecommand \natexlab [1]{#1}%
\providecommand \enquote  [1]{``#1''}%
\providecommand \bibnamefont  [1]{#1}%
\providecommand \bibfnamefont [1]{#1}%
\providecommand \citenamefont [1]{#1}%
\providecommand \href@noop [0]{\@secondoftwo}%
\providecommand \href [0]{\begingroup \@sanitize@url \@href}%
\providecommand \@href[1]{\@@startlink{#1}\@@href}%
\providecommand \@@href[1]{\endgroup#1\@@endlink}%
\providecommand \@sanitize@url [0]{\catcode `\\12\catcode `\$12\catcode
  `\&12\catcode `\#12\catcode `\^12\catcode `\_12\catcode `\%12\relax}%
\providecommand \@@startlink[1]{}%
\providecommand \@@endlink[0]{}%
\providecommand \url  [0]{\begingroup\@sanitize@url \@url }%
\providecommand \@url [1]{\endgroup\@href {#1}{\urlprefix }}%
\providecommand \urlprefix  [0]{URL }%
\providecommand \Eprint [0]{\href }%
\providecommand \doibase [0]{http://dx.doi.org/}%
\providecommand \selectlanguage [0]{\@gobble}%
\providecommand \bibinfo  [0]{\@secondoftwo}%
\providecommand \bibfield  [0]{\@secondoftwo}%
\providecommand \translation [1]{[#1]}%
\providecommand \BibitemOpen [0]{}%
\providecommand \bibitemStop [0]{}%
\providecommand \bibitemNoStop [0]{.\EOS\space}%
\providecommand \EOS [0]{\spacefactor3000\relax}%
\providecommand \BibitemShut  [1]{\csname bibitem#1\endcsname}%
\let\auto@bib@innerbib\@empty
\bibitem [{\citenamefont {Young}\ \emph {et~al.}(2012)\citenamefont {Young},
  \citenamefont {Zaheer}, \citenamefont {Teo}, \citenamefont {Kane},
  \citenamefont {Mele},\ and\ \citenamefont {Rappe}}]{young.zaheer.2012}%
  \BibitemOpen
  \bibfield  {author} {\bibinfo {author} {\bibfnamefont {S.~M.}\ \bibnamefont
  {Young}}, \bibinfo {author} {\bibfnamefont {S.}~\bibnamefont {Zaheer}},
  \bibinfo {author} {\bibfnamefont {J.~C.}\ \bibnamefont {Teo}}, \bibinfo
  {author} {\bibfnamefont {C.~L.}\ \bibnamefont {Kane}}, \bibinfo {author}
  {\bibfnamefont {E.~J.}\ \bibnamefont {Mele}}, \ and\ \bibinfo {author}
  {\bibfnamefont {A.~M.}\ \bibnamefont {Rappe}},\ }\href
  {https://doi.org/10.1103/PhysRevLett.108.140405} {\bibfield  {journal}
  {\bibinfo  {journal} {Physical Review Letters}\ }\textbf {\bibinfo {volume}
  {108}},\ \bibinfo {pages} {140405} (\bibinfo {year} {2012})}\BibitemShut
  {NoStop}%
\bibitem [{\citenamefont {Armitage}, \citenamefont {Mele},\ and\ \citenamefont
  {Vishwanath}(2018)}]{armitage.mele.2018}%
  \BibitemOpen
  \bibfield  {author} {\bibinfo {author} {\bibfnamefont {N.}~\bibnamefont
  {Armitage}}, \bibinfo {author} {\bibfnamefont {E.}~\bibnamefont {Mele}}, \
  and\ \bibinfo {author} {\bibfnamefont {A.}~\bibnamefont {Vishwanath}},\
  }\href {https://doi.org/10.1103/RevModPhys.90.015001} {\bibfield  {journal}
  {\bibinfo  {journal} {Reviews of Modern Physics}\ }\textbf {\bibinfo {volume}
  {90}},\ \bibinfo {pages} {015001} (\bibinfo {year} {2018})}\BibitemShut
  {NoStop}%
\bibitem [{\citenamefont {Wang}\ \emph {et~al.}(2012)\citenamefont {Wang},
  \citenamefont {Sun}, \citenamefont {Chen}, \citenamefont {Franchini},
  \citenamefont {Xu}, \citenamefont {Weng}, \citenamefont {Dai},\ and\
  \citenamefont {Fang}}]{wang.fang.2012}%
  \BibitemOpen
  \bibfield  {author} {\bibinfo {author} {\bibfnamefont {Z.}~\bibnamefont
  {Wang}}, \bibinfo {author} {\bibfnamefont {Y.}~\bibnamefont {Sun}}, \bibinfo
  {author} {\bibfnamefont {X.-Q.}\ \bibnamefont {Chen}}, \bibinfo {author}
  {\bibfnamefont {C.}~\bibnamefont {Franchini}}, \bibinfo {author}
  {\bibfnamefont {G.}~\bibnamefont {Xu}}, \bibinfo {author} {\bibfnamefont
  {H.}~\bibnamefont {Weng}}, \bibinfo {author} {\bibfnamefont {X.}~\bibnamefont
  {Dai}}, \ and\ \bibinfo {author} {\bibfnamefont {Z.}~\bibnamefont {Fang}},\
  }\href {https://doi.org/10.1103/PhysRevB.85.195320} {\bibfield  {journal}
  {\bibinfo  {journal} {Physical Review B}\ }\textbf {\bibinfo {volume} {85}},\
  \bibinfo {pages} {195320} (\bibinfo {year} {2012})}\BibitemShut {NoStop}%
\bibitem [{\citenamefont {Mosca~Conte}, \citenamefont {Pulci},\ and\
  \citenamefont {Bechstedt}(2017)}]{conte.pulci.2017}%
  \BibitemOpen
  \bibfield  {author} {\bibinfo {author} {\bibfnamefont {A.}~\bibnamefont
  {Mosca~Conte}}, \bibinfo {author} {\bibfnamefont {O.}~\bibnamefont {Pulci}},
  \ and\ \bibinfo {author} {\bibfnamefont {F.}~\bibnamefont {Bechstedt}},\
  }\href {https://doi.org/10.1038/srep45500} {\bibfield  {journal} {\bibinfo
  {journal} {Scientific Reports}\ }\textbf {\bibinfo {volume} {7}},\ \bibinfo
  {pages} {45500} (\bibinfo {year} {2017})}\BibitemShut {NoStop}%
\bibitem [{\citenamefont {Burkov}\ and\ \citenamefont
  {Balents}(2011)}]{burkov.balents.2011}%
  \BibitemOpen
  \bibfield  {author} {\bibinfo {author} {\bibfnamefont {A.}~\bibnamefont
  {Burkov}}\ and\ \bibinfo {author} {\bibfnamefont {L.}~\bibnamefont
  {Balents}},\ }\href {https://doi.org/10.1103/PhysRevLett.107.127205}
  {\bibfield  {journal} {\bibinfo  {journal} {Physical Review Letters}\
  }\textbf {\bibinfo {volume} {107}},\ \bibinfo {pages} {127205} (\bibinfo
  {year} {2011})}\BibitemShut {NoStop}%
\bibitem [{\citenamefont {Burkov}, \citenamefont {Hook},\ and\ \citenamefont
  {Balents}(2011)}]{burkov.hook.2011}%
  \BibitemOpen
  \bibfield  {author} {\bibinfo {author} {\bibfnamefont {A.}~\bibnamefont
  {Burkov}}, \bibinfo {author} {\bibfnamefont {M.}~\bibnamefont {Hook}}, \ and\
  \bibinfo {author} {\bibfnamefont {L.}~\bibnamefont {Balents}},\ }\href
  {https://doi.org/10.1103/PhysRevB.84.235126} {\bibfield  {journal} {\bibinfo
  {journal} {Physical Review B}\ }\textbf {\bibinfo {volume} {84}},\ \bibinfo
  {pages} {235126} (\bibinfo {year} {2011})}\BibitemShut {NoStop}%
\bibitem [{\citenamefont {Wan}\ \emph {et~al.}(2011)\citenamefont {Wan},
  \citenamefont {Turner}, \citenamefont {Vishwanath},\ and\ \citenamefont
  {Savrasov}}]{wan.turner.2011}%
  \BibitemOpen
  \bibfield  {author} {\bibinfo {author} {\bibfnamefont {X.}~\bibnamefont
  {Wan}}, \bibinfo {author} {\bibfnamefont {A.~M.}\ \bibnamefont {Turner}},
  \bibinfo {author} {\bibfnamefont {A.}~\bibnamefont {Vishwanath}}, \ and\
  \bibinfo {author} {\bibfnamefont {S.~Y.}\ \bibnamefont {Savrasov}},\
  }\href@noop {} {\bibfield  {journal} {\bibinfo  {journal} {Physical Review
  B}\ }\textbf {\bibinfo {volume} {83}},\ \bibinfo {pages} {205101} (\bibinfo
  {year} {2011})}\BibitemShut {NoStop}%
\bibitem [{\citenamefont {Chang}\ \emph {et~al.}(2013)\citenamefont {Chang},
  \citenamefont {Zhang}, \citenamefont {Feng}, \citenamefont {Shen},
  \citenamefont {Zhang}, \citenamefont {Guo}, \citenamefont {Li}, \citenamefont
  {Ou}, \citenamefont {Wei}, \citenamefont {Wang} \emph
  {et~al.}}]{chang.zhang.2013}%
  \BibitemOpen
  \bibfield  {author} {\bibinfo {author} {\bibfnamefont {C.-Z.}\ \bibnamefont
  {Chang}}, \bibinfo {author} {\bibfnamefont {J.}~\bibnamefont {Zhang}},
  \bibinfo {author} {\bibfnamefont {X.}~\bibnamefont {Feng}}, \bibinfo {author}
  {\bibfnamefont {J.}~\bibnamefont {Shen}}, \bibinfo {author} {\bibfnamefont
  {Z.}~\bibnamefont {Zhang}}, \bibinfo {author} {\bibfnamefont
  {M.}~\bibnamefont {Guo}}, \bibinfo {author} {\bibfnamefont {K.}~\bibnamefont
  {Li}}, \bibinfo {author} {\bibfnamefont {Y.}~\bibnamefont {Ou}}, \bibinfo
  {author} {\bibfnamefont {P.}~\bibnamefont {Wei}}, \bibinfo {author}
  {\bibfnamefont {L.-L.}\ \bibnamefont {Wang}},  \emph {et~al.},\ }\href
  {https://doi.org/10.1126/science.1234414} {\bibfield  {journal} {\bibinfo
  {journal} {Science}\ ,\ \bibinfo {pages} {1232003}} (\bibinfo {year}
  {2013})}\BibitemShut {NoStop}%
\bibitem [{\citenamefont {Yu}\ \emph {et~al.}(2010)\citenamefont {Yu},
  \citenamefont {Zhang}, \citenamefont {Zhang}, \citenamefont {Zhang},
  \citenamefont {Dai},\ and\ \citenamefont {Fang}}]{yu.zhang.2010}%
  \BibitemOpen
  \bibfield  {author} {\bibinfo {author} {\bibfnamefont {R.}~\bibnamefont
  {Yu}}, \bibinfo {author} {\bibfnamefont {W.}~\bibnamefont {Zhang}}, \bibinfo
  {author} {\bibfnamefont {H.-J.}\ \bibnamefont {Zhang}}, \bibinfo {author}
  {\bibfnamefont {S.-C.}\ \bibnamefont {Zhang}}, \bibinfo {author}
  {\bibfnamefont {X.}~\bibnamefont {Dai}}, \ and\ \bibinfo {author}
  {\bibfnamefont {Z.}~\bibnamefont {Fang}},\ }\href
  {https://doi.org/10.1126/science.1187485} {\bibfield  {journal} {\bibinfo
  {journal} {Science}\ }\textbf {\bibinfo {volume} {329}},\ \bibinfo {pages}
  {61} (\bibinfo {year} {2010})}\BibitemShut {NoStop}%
\bibitem [{\citenamefont {Murakami}, \citenamefont {Nagaosa},\ and\
  \citenamefont {Zhang}(2003)}]{murakami.nagosa.2003}%
  \BibitemOpen
  \bibfield  {author} {\bibinfo {author} {\bibfnamefont {S.}~\bibnamefont
  {Murakami}}, \bibinfo {author} {\bibfnamefont {N.}~\bibnamefont {Nagaosa}}, \
  and\ \bibinfo {author} {\bibfnamefont {S.-C.}\ \bibnamefont {Zhang}},\ }\href
  {https://doi.org/10.1126/science.1087128} {\bibfield  {journal} {\bibinfo
  {journal} {Science}\ }\textbf {\bibinfo {volume} {301}},\ \bibinfo {pages}
  {1348} (\bibinfo {year} {2003})}\BibitemShut {NoStop}%
\bibitem [{\citenamefont {Sinova}\ \emph {et~al.}(2004)\citenamefont {Sinova},
  \citenamefont {Culcer}, \citenamefont {Niu}, \citenamefont {Sinitsyn},
  \citenamefont {Jungwirth},\ and\ \citenamefont
  {MacDonald}}]{sinova.culcer.2004}%
  \BibitemOpen
  \bibfield  {author} {\bibinfo {author} {\bibfnamefont {J.}~\bibnamefont
  {Sinova}}, \bibinfo {author} {\bibfnamefont {D.}~\bibnamefont {Culcer}},
  \bibinfo {author} {\bibfnamefont {Q.}~\bibnamefont {Niu}}, \bibinfo {author}
  {\bibfnamefont {N.}~\bibnamefont {Sinitsyn}}, \bibinfo {author}
  {\bibfnamefont {T.}~\bibnamefont {Jungwirth}}, \ and\ \bibinfo {author}
  {\bibfnamefont {A.}~\bibnamefont {MacDonald}},\ }\href
  {https://doi.org/10.1103/PhysRevLett.92.126603} {\bibfield  {journal}
  {\bibinfo  {journal} {Physical Review Letters}\ }\textbf {\bibinfo {volume}
  {92}},\ \bibinfo {pages} {126603} (\bibinfo {year} {2004})}\BibitemShut
  {NoStop}%
\bibitem [{\citenamefont {Rauch}\ \emph {et~al.}(2017)\citenamefont {Rauch},
  \citenamefont {Minh}, \citenamefont {Henk},\ and\ \citenamefont
  {Mertig}}]{rauch.mertig.2017}%
  \BibitemOpen
  \bibfield  {author} {\bibinfo {author} {\bibfnamefont {T.}~\bibnamefont
  {Rauch}}, \bibinfo {author} {\bibfnamefont {H.~N.}\ \bibnamefont {Minh}},
  \bibinfo {author} {\bibfnamefont {J.}~\bibnamefont {Henk}}, \ and\ \bibinfo
  {author} {\bibfnamefont {I.}~\bibnamefont {Mertig}},\ }\href
  {https://doi.org/10.1103/PhysRevB.96.235103} {\bibfield  {journal} {\bibinfo
  {journal} {Physical Review B}\ }\textbf {\bibinfo {volume} {96}},\ \bibinfo
  {pages} {235103} (\bibinfo {year} {2017})}\BibitemShut {NoStop}%
\bibitem [{\citenamefont {Huang}\ \emph
  {et~al.}(2015{\natexlab{a}})\citenamefont {Huang}, \citenamefont {Zhao},
  \citenamefont {Long}, \citenamefont {Wang}, \citenamefont {Chen},
  \citenamefont {Yang}, \citenamefont {Liang}, \citenamefont {Xue},
  \citenamefont {Weng}, \citenamefont {Fang} \emph {et~al.}}]{huang.zhao.2015}%
  \BibitemOpen
  \bibfield  {author} {\bibinfo {author} {\bibfnamefont {X.}~\bibnamefont
  {Huang}}, \bibinfo {author} {\bibfnamefont {L.}~\bibnamefont {Zhao}},
  \bibinfo {author} {\bibfnamefont {Y.}~\bibnamefont {Long}}, \bibinfo {author}
  {\bibfnamefont {P.}~\bibnamefont {Wang}}, \bibinfo {author} {\bibfnamefont
  {D.}~\bibnamefont {Chen}}, \bibinfo {author} {\bibfnamefont {Z.}~\bibnamefont
  {Yang}}, \bibinfo {author} {\bibfnamefont {H.}~\bibnamefont {Liang}},
  \bibinfo {author} {\bibfnamefont {M.}~\bibnamefont {Xue}}, \bibinfo {author}
  {\bibfnamefont {H.}~\bibnamefont {Weng}}, \bibinfo {author} {\bibfnamefont
  {Z.}~\bibnamefont {Fang}},  \emph {et~al.},\ }\href
  {https://doi.org/10.1103/PhysRevX.5.031023} {\bibfield  {journal} {\bibinfo
  {journal} {Physical Review X}\ }\textbf {\bibinfo {volume} {5}},\ \bibinfo
  {pages} {031023} (\bibinfo {year} {2015}{\natexlab{a}})}\BibitemShut
  {NoStop}%
\bibitem [{\citenamefont {Nielsen}\ and\ \citenamefont
  {Ninomiya}(1983)}]{nielsen.nunomiya.1983}%
  \BibitemOpen
  \bibfield  {author} {\bibinfo {author} {\bibfnamefont {H.~B.}\ \bibnamefont
  {Nielsen}}\ and\ \bibinfo {author} {\bibfnamefont {M.}~\bibnamefont
  {Ninomiya}},\ }\href {https://doi.org/10.1016/0370-2693(83)91529-0}
  {\bibfield  {journal} {\bibinfo  {journal} {Physics Letters B}\ }\textbf
  {\bibinfo {volume} {130}},\ \bibinfo {pages} {389} (\bibinfo {year}
  {1983})}\BibitemShut {NoStop}%
\bibitem [{\citenamefont {Hills}, \citenamefont {Kusmartseva},\ and\
  \citenamefont {Kusmartsev}(2017)}]{hills.kusmartseva.2017}%
  \BibitemOpen
  \bibfield  {author} {\bibinfo {author} {\bibfnamefont {R.~D.}\ \bibnamefont
  {Hills}}, \bibinfo {author} {\bibfnamefont {A.}~\bibnamefont {Kusmartseva}},
  \ and\ \bibinfo {author} {\bibfnamefont {F.}~\bibnamefont {Kusmartsev}},\
  }\href {https://doi.org/10.1103/PhysRevB.95.214103} {\bibfield  {journal}
  {\bibinfo  {journal} {Physical Review B}\ }\textbf {\bibinfo {volume} {95}},\
  \bibinfo {pages} {214103} (\bibinfo {year} {2017})}\BibitemShut {NoStop}%
\bibitem [{\citenamefont {Huang}\ \emph
  {et~al.}(2015{\natexlab{b}})\citenamefont {Huang}, \citenamefont {Xu},
  \citenamefont {Belopolski}, \citenamefont {Lee}, \citenamefont {Chang},
  \citenamefont {Wang}, \citenamefont {Alidoust}, \citenamefont {Bian},
  \citenamefont {Neupane}, \citenamefont {Zhang} \emph
  {et~al.}}]{huang.xu.2015}%
  \BibitemOpen
  \bibfield  {author} {\bibinfo {author} {\bibfnamefont {S.-M.}\ \bibnamefont
  {Huang}}, \bibinfo {author} {\bibfnamefont {S.-Y.}\ \bibnamefont {Xu}},
  \bibinfo {author} {\bibfnamefont {I.}~\bibnamefont {Belopolski}}, \bibinfo
  {author} {\bibfnamefont {C.-C.}\ \bibnamefont {Lee}}, \bibinfo {author}
  {\bibfnamefont {G.}~\bibnamefont {Chang}}, \bibinfo {author} {\bibfnamefont
  {B.}~\bibnamefont {Wang}}, \bibinfo {author} {\bibfnamefont {N.}~\bibnamefont
  {Alidoust}}, \bibinfo {author} {\bibfnamefont {G.}~\bibnamefont {Bian}},
  \bibinfo {author} {\bibfnamefont {M.}~\bibnamefont {Neupane}}, \bibinfo
  {author} {\bibfnamefont {C.}~\bibnamefont {Zhang}},  \emph {et~al.},\ }\href
  {http://dx.doi.org/10.1038%2Fncomms8373} {\bibfield  {journal} {\bibinfo
  {journal} {Nature Communications}\ }\textbf {\bibinfo {volume} {6}},\
  \bibinfo {pages} {7373} (\bibinfo {year} {2015}{\natexlab{b}})}\BibitemShut
  {NoStop}%
\bibitem [{\citenamefont {Li}\ \emph {et~al.}(2016)\citenamefont {Li},
  \citenamefont {Chen}, \citenamefont {Jin}, \citenamefont {Gan}, \citenamefont
  {Wang}, \citenamefont {Guo},\ and\ \citenamefont {Chen}}]{li.2016}%
  \BibitemOpen
  \bibfield  {author} {\bibinfo {author} {\bibfnamefont {Z.}~\bibnamefont
  {Li}}, \bibinfo {author} {\bibfnamefont {H.}~\bibnamefont {Chen}}, \bibinfo
  {author} {\bibfnamefont {S.}~\bibnamefont {Jin}}, \bibinfo {author}
  {\bibfnamefont {D.}~\bibnamefont {Gan}}, \bibinfo {author} {\bibfnamefont
  {W.}~\bibnamefont {Wang}}, \bibinfo {author} {\bibfnamefont {L.}~\bibnamefont
  {Guo}}, \ and\ \bibinfo {author} {\bibfnamefont {X.}~\bibnamefont {Chen}},\
  }\href@noop {} {\bibfield  {journal} {\bibinfo  {journal} {Crystal Growth \&
  Design}\ }\textbf {\bibinfo {volume} {16}},\ \bibinfo {pages} {1172}
  (\bibinfo {year} {2016})}\BibitemShut {NoStop}%
\bibitem [{\citenamefont {Weng}\ \emph {et~al.}(2015)\citenamefont {Weng},
  \citenamefont {Fang}, \citenamefont {Fang}, \citenamefont {Bernevig},\ and\
  \citenamefont {Dai}}]{weng.fang.2015}%
  \BibitemOpen
  \bibfield  {author} {\bibinfo {author} {\bibfnamefont {H.}~\bibnamefont
  {Weng}}, \bibinfo {author} {\bibfnamefont {C.}~\bibnamefont {Fang}}, \bibinfo
  {author} {\bibfnamefont {Z.}~\bibnamefont {Fang}}, \bibinfo {author}
  {\bibfnamefont {B.~A.}\ \bibnamefont {Bernevig}}, \ and\ \bibinfo {author}
  {\bibfnamefont {X.}~\bibnamefont {Dai}},\ }\href
  {https://doi.org/10.1103/PhysRevX.5.011029} {\bibfield  {journal} {\bibinfo
  {journal} {Physical Review X}\ }\textbf {\bibinfo {volume} {5}},\ \bibinfo
  {pages} {011029} (\bibinfo {year} {2015})}\BibitemShut {NoStop}%
\bibitem [{\citenamefont {Sun}, \citenamefont {Wu},\ and\ \citenamefont
  {Yan}(2015)}]{sun.wu.2015}%
  \BibitemOpen
  \bibfield  {author} {\bibinfo {author} {\bibfnamefont {Y.}~\bibnamefont
  {Sun}}, \bibinfo {author} {\bibfnamefont {S.-C.}\ \bibnamefont {Wu}}, \ and\
  \bibinfo {author} {\bibfnamefont {B.}~\bibnamefont {Yan}},\ }\href
  {https://doi.org/10.1103/PhysRevB.92.115428} {\bibfield  {journal} {\bibinfo
  {journal} {Physical Review B}\ }\textbf {\bibinfo {volume} {92}},\ \bibinfo
  {pages} {115428} (\bibinfo {year} {2015})}\BibitemShut {NoStop}%
\bibitem [{\citenamefont {Lee}\ \emph {et~al.}(2015)\citenamefont {Lee},
  \citenamefont {Xu}, \citenamefont {Huang}, \citenamefont {Sanchez},
  \citenamefont {Belopolski}, \citenamefont {Chang}, \citenamefont {Bian},
  \citenamefont {Alidoust}, \citenamefont {Zheng}, \citenamefont {Neupane}
  \emph {et~al.}}]{lee.xu.2015}%
  \BibitemOpen
  \bibfield  {author} {\bibinfo {author} {\bibfnamefont {C.-C.}\ \bibnamefont
  {Lee}}, \bibinfo {author} {\bibfnamefont {S.-Y.}\ \bibnamefont {Xu}},
  \bibinfo {author} {\bibfnamefont {S.-M.}\ \bibnamefont {Huang}}, \bibinfo
  {author} {\bibfnamefont {D.~S.}\ \bibnamefont {Sanchez}}, \bibinfo {author}
  {\bibfnamefont {I.}~\bibnamefont {Belopolski}}, \bibinfo {author}
  {\bibfnamefont {G.}~\bibnamefont {Chang}}, \bibinfo {author} {\bibfnamefont
  {G.}~\bibnamefont {Bian}}, \bibinfo {author} {\bibfnamefont {N.}~\bibnamefont
  {Alidoust}}, \bibinfo {author} {\bibfnamefont {H.}~\bibnamefont {Zheng}},
  \bibinfo {author} {\bibfnamefont {M.}~\bibnamefont {Neupane}},  \emph
  {et~al.},\ }\href {https://doi.org/10.1103/PhysRevB.92.235104} {\bibfield
  {journal} {\bibinfo  {journal} {Physical Review B}\ }\textbf {\bibinfo
  {volume} {92}},\ \bibinfo {pages} {235104} (\bibinfo {year}
  {2015})}\BibitemShut {NoStop}%
\bibitem [{\citenamefont {Grassano}\ \emph {et~al.}(2018)\citenamefont
  {Grassano}, \citenamefont {Pulci}, \citenamefont {Mosca~Conte},\ and\
  \citenamefont {Bechstedt}}]{grassano.pulci.2018}%
  \BibitemOpen
  \bibfield  {author} {\bibinfo {author} {\bibfnamefont {D.}~\bibnamefont
  {Grassano}}, \bibinfo {author} {\bibfnamefont {O.}~\bibnamefont {Pulci}},
  \bibinfo {author} {\bibfnamefont {A.}~\bibnamefont {Mosca~Conte}}, \ and\
  \bibinfo {author} {\bibfnamefont {F.}~\bibnamefont {Bechstedt}},\ }\href
  {https://doi.org/10.1038/s41598-018-21465-z} {\bibfield  {journal} {\bibinfo
  {journal} {Scientific Reports}\ }\textbf {\bibinfo {volume} {8}},\ \bibinfo
  {pages} {3534} (\bibinfo {year} {2018})}\BibitemShut {NoStop}%
\bibitem [{\citenamefont {Xu}\ \emph {et~al.}(2015{\natexlab{a}})\citenamefont
  {Xu}, \citenamefont {Belopolski}, \citenamefont {Alidoust}, \citenamefont
  {Neupane}, \citenamefont {Bian}, \citenamefont {Zhang}, \citenamefont
  {Sankar}, \citenamefont {Chang}, \citenamefont {Yuan}, \citenamefont {Lee}
  \emph {et~al.}}]{xu.belopolski.2015.TaAs}%
  \BibitemOpen
  \bibfield  {author} {\bibinfo {author} {\bibfnamefont {S.-Y.}\ \bibnamefont
  {Xu}}, \bibinfo {author} {\bibfnamefont {I.}~\bibnamefont {Belopolski}},
  \bibinfo {author} {\bibfnamefont {N.}~\bibnamefont {Alidoust}}, \bibinfo
  {author} {\bibfnamefont {M.}~\bibnamefont {Neupane}}, \bibinfo {author}
  {\bibfnamefont {G.}~\bibnamefont {Bian}}, \bibinfo {author} {\bibfnamefont
  {C.}~\bibnamefont {Zhang}}, \bibinfo {author} {\bibfnamefont
  {R.}~\bibnamefont {Sankar}}, \bibinfo {author} {\bibfnamefont
  {G.}~\bibnamefont {Chang}}, \bibinfo {author} {\bibfnamefont
  {Z.}~\bibnamefont {Yuan}}, \bibinfo {author} {\bibfnamefont {C.-C.}\
  \bibnamefont {Lee}},  \emph {et~al.},\ }\href@noop {} {\bibfield  {journal}
  {\bibinfo  {journal} {Science}\ }\textbf {\bibinfo {volume} {349}},\ \bibinfo
  {pages} {613} (\bibinfo {year} {2015}{\natexlab{a}})}\BibitemShut {NoStop}%
\bibitem [{\citenamefont {Lv}\ \emph {et~al.}(2015)\citenamefont {Lv},
  \citenamefont {Weng}, \citenamefont {Fu}, \citenamefont {Wang}, \citenamefont
  {Miao}, \citenamefont {Ma}, \citenamefont {Richard}, \citenamefont {Huang},
  \citenamefont {Zhao}, \citenamefont {Chen} \emph {et~al.}}]{lv.weng.2015}%
  \BibitemOpen
  \bibfield  {author} {\bibinfo {author} {\bibfnamefont {B.}~\bibnamefont
  {Lv}}, \bibinfo {author} {\bibfnamefont {H.}~\bibnamefont {Weng}}, \bibinfo
  {author} {\bibfnamefont {B.}~\bibnamefont {Fu}}, \bibinfo {author}
  {\bibfnamefont {X.}~\bibnamefont {Wang}}, \bibinfo {author} {\bibfnamefont
  {H.}~\bibnamefont {Miao}}, \bibinfo {author} {\bibfnamefont {J.}~\bibnamefont
  {Ma}}, \bibinfo {author} {\bibfnamefont {P.}~\bibnamefont {Richard}},
  \bibinfo {author} {\bibfnamefont {X.}~\bibnamefont {Huang}}, \bibinfo
  {author} {\bibfnamefont {L.}~\bibnamefont {Zhao}}, \bibinfo {author}
  {\bibfnamefont {G.}~\bibnamefont {Chen}},  \emph {et~al.},\ }\href
  {https://doi.org/10.1103/PhysRevX.5.031013} {\bibfield  {journal} {\bibinfo
  {journal} {Physical Review X}\ }\textbf {\bibinfo {volume} {5}},\ \bibinfo
  {pages} {031013} (\bibinfo {year} {2015})}\BibitemShut {NoStop}%
\bibitem [{\citenamefont {Xu}\ \emph {et~al.}(2015{\natexlab{b}})\citenamefont
  {Xu}, \citenamefont {Belopolski}, \citenamefont {Sanchez}, \citenamefont
  {Zhang}, \citenamefont {Chang}, \citenamefont {Guo}, \citenamefont {Bian},
  \citenamefont {Yuan}, \citenamefont {Lu}, \citenamefont {Chang} \emph
  {et~al.}}]{xu.belopolski.2015.TaP}%
  \BibitemOpen
  \bibfield  {author} {\bibinfo {author} {\bibfnamefont {S.-Y.}\ \bibnamefont
  {Xu}}, \bibinfo {author} {\bibfnamefont {I.}~\bibnamefont {Belopolski}},
  \bibinfo {author} {\bibfnamefont {D.~S.}\ \bibnamefont {Sanchez}}, \bibinfo
  {author} {\bibfnamefont {C.}~\bibnamefont {Zhang}}, \bibinfo {author}
  {\bibfnamefont {G.}~\bibnamefont {Chang}}, \bibinfo {author} {\bibfnamefont
  {C.}~\bibnamefont {Guo}}, \bibinfo {author} {\bibfnamefont {G.}~\bibnamefont
  {Bian}}, \bibinfo {author} {\bibfnamefont {Z.}~\bibnamefont {Yuan}}, \bibinfo
  {author} {\bibfnamefont {H.}~\bibnamefont {Lu}}, \bibinfo {author}
  {\bibfnamefont {T.-R.}\ \bibnamefont {Chang}},  \emph {et~al.},\ }\href
  {https://dx.doi.org/10.1126%2Fsciadv.1501092} {\bibfield  {journal} {\bibinfo
   {journal} {Science Advances}\ }\textbf {\bibinfo {volume} {1}},\ \bibinfo
  {pages} {e1501092} (\bibinfo {year} {2015}{\natexlab{b}})}\BibitemShut
  {NoStop}%
\bibitem [{\citenamefont {Xu}\ \emph {et~al.}(2015{\natexlab{c}})\citenamefont
  {Xu}, \citenamefont {Alidoust}, \citenamefont {Belopolski}, \citenamefont
  {Yuan}, \citenamefont {Bian}, \citenamefont {Chang}, \citenamefont {Zheng},
  \citenamefont {Strocov}, \citenamefont {Sanchez}, \citenamefont {Chang} \emph
  {et~al.}}]{xu.alidoust.2015}%
  \BibitemOpen
  \bibfield  {author} {\bibinfo {author} {\bibfnamefont {S.-Y.}\ \bibnamefont
  {Xu}}, \bibinfo {author} {\bibfnamefont {N.}~\bibnamefont {Alidoust}},
  \bibinfo {author} {\bibfnamefont {I.}~\bibnamefont {Belopolski}}, \bibinfo
  {author} {\bibfnamefont {Z.}~\bibnamefont {Yuan}}, \bibinfo {author}
  {\bibfnamefont {G.}~\bibnamefont {Bian}}, \bibinfo {author} {\bibfnamefont
  {T.-R.}\ \bibnamefont {Chang}}, \bibinfo {author} {\bibfnamefont
  {H.}~\bibnamefont {Zheng}}, \bibinfo {author} {\bibfnamefont {V.~N.}\
  \bibnamefont {Strocov}}, \bibinfo {author} {\bibfnamefont {D.~S.}\
  \bibnamefont {Sanchez}}, \bibinfo {author} {\bibfnamefont {G.}~\bibnamefont
  {Chang}},  \emph {et~al.},\ }\href {https://doi.org/10.1038/nphys3437}
  {\bibfield  {journal} {\bibinfo  {journal} {Nature Physics}\ }\textbf
  {\bibinfo {volume} {11}},\ \bibinfo {pages} {748} (\bibinfo {year}
  {2015}{\natexlab{c}})}\BibitemShut {NoStop}%
\bibitem [{\citenamefont {Souma}\ \emph {et~al.}(2016)\citenamefont {Souma},
  \citenamefont {Wang}, \citenamefont {Kotaka}, \citenamefont {Sato},
  \citenamefont {Nakayama}, \citenamefont {Tanaka}, \citenamefont {Kimizuka},
  \citenamefont {Takahashi}, \citenamefont {Yamauchi}, \citenamefont {Oguchi}
  \emph {et~al.}}]{souma.wang.2016}%
  \BibitemOpen
  \bibfield  {author} {\bibinfo {author} {\bibfnamefont {S.}~\bibnamefont
  {Souma}}, \bibinfo {author} {\bibfnamefont {Z.}~\bibnamefont {Wang}},
  \bibinfo {author} {\bibfnamefont {H.}~\bibnamefont {Kotaka}}, \bibinfo
  {author} {\bibfnamefont {T.}~\bibnamefont {Sato}}, \bibinfo {author}
  {\bibfnamefont {K.}~\bibnamefont {Nakayama}}, \bibinfo {author}
  {\bibfnamefont {Y.}~\bibnamefont {Tanaka}}, \bibinfo {author} {\bibfnamefont
  {H.}~\bibnamefont {Kimizuka}}, \bibinfo {author} {\bibfnamefont
  {T.}~\bibnamefont {Takahashi}}, \bibinfo {author} {\bibfnamefont
  {K.}~\bibnamefont {Yamauchi}}, \bibinfo {author} {\bibfnamefont
  {T.}~\bibnamefont {Oguchi}},  \emph {et~al.},\ }\href@noop {} {\bibfield
  {journal} {\bibinfo  {journal} {Physical Review B}\ }\textbf {\bibinfo
  {volume} {93}},\ \bibinfo {pages} {161112} (\bibinfo {year}
  {2016})}\BibitemShut {NoStop}%
\bibitem [{\citenamefont {Xu}\ \emph {et~al.}(2016)\citenamefont {Xu},
  \citenamefont {Dai}, \citenamefont {Zhao}, \citenamefont {Wang},
  \citenamefont {Yang}, \citenamefont {Zhang}, \citenamefont {Liu},
  \citenamefont {Xiao}, \citenamefont {Chen}, \citenamefont {Taylor} \emph
  {et~al.}}]{xu.dai.2016}%
  \BibitemOpen
  \bibfield  {author} {\bibinfo {author} {\bibfnamefont {B.}~\bibnamefont
  {Xu}}, \bibinfo {author} {\bibfnamefont {Y.}~\bibnamefont {Dai}}, \bibinfo
  {author} {\bibfnamefont {L.}~\bibnamefont {Zhao}}, \bibinfo {author}
  {\bibfnamefont {K.}~\bibnamefont {Wang}}, \bibinfo {author} {\bibfnamefont
  {R.}~\bibnamefont {Yang}}, \bibinfo {author} {\bibfnamefont {W.}~\bibnamefont
  {Zhang}}, \bibinfo {author} {\bibfnamefont {J.}~\bibnamefont {Liu}}, \bibinfo
  {author} {\bibfnamefont {H.}~\bibnamefont {Xiao}}, \bibinfo {author}
  {\bibfnamefont {G.}~\bibnamefont {Chen}}, \bibinfo {author} {\bibfnamefont
  {A.}~\bibnamefont {Taylor}},  \emph {et~al.},\ }\href
  {https://doi.org/10.1103/PhysRevB.93.121110} {\bibfield  {journal} {\bibinfo
  {journal} {Physical Review B}\ }\textbf {\bibinfo {volume} {93}},\ \bibinfo
  {pages} {121110} (\bibinfo {year} {2016})}\BibitemShut {NoStop}%
\bibitem [{\citenamefont {Kimura}\ \emph {et~al.}(2017)\citenamefont {Kimura},
  \citenamefont {Yokoyama}, \citenamefont {Watanabe}, \citenamefont
  {Sichelschmidt}, \citenamefont {S{\"u}{\ss}}, \citenamefont {Schmidt},\ and\
  \citenamefont {Felser}}]{kimura2017optical}%
  \BibitemOpen
  \bibfield  {author} {\bibinfo {author} {\bibfnamefont {S.-i.}\ \bibnamefont
  {Kimura}}, \bibinfo {author} {\bibfnamefont {H.}~\bibnamefont {Yokoyama}},
  \bibinfo {author} {\bibfnamefont {H.}~\bibnamefont {Watanabe}}, \bibinfo
  {author} {\bibfnamefont {J.}~\bibnamefont {Sichelschmidt}}, \bibinfo {author}
  {\bibfnamefont {V.}~\bibnamefont {S{\"u}{\ss}}}, \bibinfo {author}
  {\bibfnamefont {M.}~\bibnamefont {Schmidt}}, \ and\ \bibinfo {author}
  {\bibfnamefont {C.}~\bibnamefont {Felser}},\ }\href@noop {} {\bibfield
  {journal} {\bibinfo  {journal} {Physical Review B}\ }\textbf {\bibinfo
  {volume} {96}},\ \bibinfo {pages} {075119} (\bibinfo {year}
  {2017})}\BibitemShut {NoStop}%
\bibitem [{\citenamefont {Giannozzi}\ \emph {et~al.}(2009)\citenamefont
  {Giannozzi}, \citenamefont {Baroni}, \citenamefont {Bonini}, \citenamefont
  {Calandra}, \citenamefont {Car}, \citenamefont {Cavazzoni}, \citenamefont
  {Ceresoli}, \citenamefont {Chiarotti}, \citenamefont {Cococcioni},
  \citenamefont {Dabo} \emph {et~al.}}]{espresso.2009}%
  \BibitemOpen
  \bibfield  {author} {\bibinfo {author} {\bibfnamefont {P.}~\bibnamefont
  {Giannozzi}}, \bibinfo {author} {\bibfnamefont {S.}~\bibnamefont {Baroni}},
  \bibinfo {author} {\bibfnamefont {N.}~\bibnamefont {Bonini}}, \bibinfo
  {author} {\bibfnamefont {M.}~\bibnamefont {Calandra}}, \bibinfo {author}
  {\bibfnamefont {R.}~\bibnamefont {Car}}, \bibinfo {author} {\bibfnamefont
  {C.}~\bibnamefont {Cavazzoni}}, \bibinfo {author} {\bibfnamefont
  {D.}~\bibnamefont {Ceresoli}}, \bibinfo {author} {\bibfnamefont {G.~L.}\
  \bibnamefont {Chiarotti}}, \bibinfo {author} {\bibfnamefont {M.}~\bibnamefont
  {Cococcioni}}, \bibinfo {author} {\bibfnamefont {I.}~\bibnamefont {Dabo}},
  \emph {et~al.},\ }\href {https://doi.org/10.1088/0953-8984/21/39/395502}
  {\bibfield  {journal} {\bibinfo  {journal} {Journal of Physics: Condensed
  matter}\ }\textbf {\bibinfo {volume} {21}},\ \bibinfo {pages} {395502}
  (\bibinfo {year} {2009})}\BibitemShut {NoStop}%
\bibitem [{\citenamefont {Giannozzi}\ \emph {et~al.}(2017)\citenamefont
  {Giannozzi}, \citenamefont {Andreussi}, \citenamefont {Brumme}, \citenamefont
  {Bunau}, \citenamefont {Nardelli}, \citenamefont {Calandra}, \citenamefont
  {Car}, \citenamefont {Cavazzoni}, \citenamefont {Ceresoli}, \citenamefont
  {Cococcioni} \emph {et~al.}}]{espresso.2017}%
  \BibitemOpen
  \bibfield  {author} {\bibinfo {author} {\bibfnamefont {P.}~\bibnamefont
  {Giannozzi}}, \bibinfo {author} {\bibfnamefont {O.}~\bibnamefont
  {Andreussi}}, \bibinfo {author} {\bibfnamefont {T.}~\bibnamefont {Brumme}},
  \bibinfo {author} {\bibfnamefont {O.}~\bibnamefont {Bunau}}, \bibinfo
  {author} {\bibfnamefont {M.~B.}\ \bibnamefont {Nardelli}}, \bibinfo {author}
  {\bibfnamefont {M.}~\bibnamefont {Calandra}}, \bibinfo {author}
  {\bibfnamefont {R.}~\bibnamefont {Car}}, \bibinfo {author} {\bibfnamefont
  {C.}~\bibnamefont {Cavazzoni}}, \bibinfo {author} {\bibfnamefont
  {D.}~\bibnamefont {Ceresoli}}, \bibinfo {author} {\bibfnamefont
  {M.}~\bibnamefont {Cococcioni}},  \emph {et~al.},\ }\href
  {https://doi.org/10.1088/1361-648X/aa8f79} {\bibfield  {journal} {\bibinfo
  {journal} {Journal of Physics: Condensed Matter}\ }\textbf {\bibinfo {volume}
  {29}},\ \bibinfo {pages} {465901} (\bibinfo {year} {2017})}\BibitemShut
  {NoStop}%
\bibitem [{\citenamefont {Perdew}, \citenamefont {Burke},\ and\ \citenamefont
  {Ernzerhof}(1996)}]{PBE.1996}%
  \BibitemOpen
  \bibfield  {author} {\bibinfo {author} {\bibfnamefont {J.~P.}\ \bibnamefont
  {Perdew}}, \bibinfo {author} {\bibfnamefont {K.}~\bibnamefont {Burke}}, \
  and\ \bibinfo {author} {\bibfnamefont {M.}~\bibnamefont {Ernzerhof}},\ }\href
  {https://doi.org/10.1103/PhysRevLett.77.3865} {\bibfield  {journal} {\bibinfo
   {journal} {Physical Review letters}\ }\textbf {\bibinfo {volume} {77}},\
  \bibinfo {pages} {3865} (\bibinfo {year} {1996})}\BibitemShut {NoStop}%
\bibitem [{\citenamefont {Hamann}(2013)}]{hamann.2013}%
  \BibitemOpen
  \bibfield  {author} {\bibinfo {author} {\bibfnamefont {D.}~\bibnamefont
  {Hamann}},\ }\href {https://doi.org/10.1103/PhysRevB.88.085117} {\bibfield
  {journal} {\bibinfo  {journal} {Physical Review B}\ }\textbf {\bibinfo
  {volume} {88}},\ \bibinfo {pages} {085117} (\bibinfo {year}
  {2013})}\BibitemShut {NoStop}%
\bibitem [{\citenamefont {Gonze}, \citenamefont {K{\"a}ckell},\ and\
  \citenamefont {Scheffler}(1990)}]{gonze.kackell.1990}%
  \BibitemOpen
  \bibfield  {author} {\bibinfo {author} {\bibfnamefont {X.}~\bibnamefont
  {Gonze}}, \bibinfo {author} {\bibfnamefont {P.}~\bibnamefont {K{\"a}ckell}},
  \ and\ \bibinfo {author} {\bibfnamefont {M.}~\bibnamefont {Scheffler}},\
  }\href {https://doi.org/10.1103/PhysRevB.41.12264} {\bibfield  {journal}
  {\bibinfo  {journal} {Physical Review B}\ }\textbf {\bibinfo {volume} {41}},\
  \bibinfo {pages} {12264} (\bibinfo {year} {1990})}\BibitemShut {NoStop}%
\bibitem [{\citenamefont {Dal~Corso}\ and\ \citenamefont
  {Mosca~Conte}(2005)}]{corso.conte.2005}%
  \BibitemOpen
  \bibfield  {author} {\bibinfo {author} {\bibfnamefont {A.}~\bibnamefont
  {Dal~Corso}}\ and\ \bibinfo {author} {\bibfnamefont {A.}~\bibnamefont
  {Mosca~Conte}},\ }\href {https://doi.org/10.1103/PhysRevB.71.115106}
  {\bibfield  {journal} {\bibinfo  {journal} {Physical Review B}\ }\textbf
  {\bibinfo {volume} {71}},\ \bibinfo {pages} {115106} (\bibinfo {year}
  {2005})}\BibitemShut {NoStop}%
\bibitem [{\citenamefont {Mosca~Conte}, \citenamefont {Fabris},\ and\
  \citenamefont {Baroni}(2008)}]{conte.fabris.2008}%
  \BibitemOpen
  \bibfield  {author} {\bibinfo {author} {\bibfnamefont {A.}~\bibnamefont
  {Mosca~Conte}}, \bibinfo {author} {\bibfnamefont {S.}~\bibnamefont {Fabris}},
  \ and\ \bibinfo {author} {\bibfnamefont {S.}~\bibnamefont {Baroni}},\ }\href
  {https://doi.org/10.1103/PhysRevB.78.014416} {\bibfield  {journal} {\bibinfo
  {journal} {Physical Review B}\ }\textbf {\bibinfo {volume} {78}},\ \bibinfo
  {pages} {014416} (\bibinfo {year} {2008})}\BibitemShut {NoStop}%
\bibitem [{\citenamefont {Marini}\ \emph {et~al.}(2009)\citenamefont {Marini},
  \citenamefont {Hogan}, \citenamefont {Gr{\"u}ning},\ and\ \citenamefont
  {Varsano}}]{yambo.2009}%
  \BibitemOpen
  \bibfield  {author} {\bibinfo {author} {\bibfnamefont {A.}~\bibnamefont
  {Marini}}, \bibinfo {author} {\bibfnamefont {C.}~\bibnamefont {Hogan}},
  \bibinfo {author} {\bibfnamefont {M.}~\bibnamefont {Gr{\"u}ning}}, \ and\
  \bibinfo {author} {\bibfnamefont {D.}~\bibnamefont {Varsano}},\ }\href
  {https://doi.org/10.1016/j.cpc.2009.02.003} {\bibfield  {journal} {\bibinfo
  {journal} {Computer Physics Communications}\ }\textbf {\bibinfo {volume}
  {180}},\ \bibinfo {pages} {1392} (\bibinfo {year} {2009})}\BibitemShut
  {NoStop}%
\bibitem [{\citenamefont {Del~Sole}\ and\ \citenamefont
  {Girlanda}(1993)}]{delsole.girlanda.1993}%
  \BibitemOpen
  \bibfield  {author} {\bibinfo {author} {\bibfnamefont {R.}~\bibnamefont
  {Del~Sole}}\ and\ \bibinfo {author} {\bibfnamefont {R.}~\bibnamefont
  {Girlanda}},\ }\href {https://doi.org/10.1103/PhysRevB.48.11789} {\bibfield
  {journal} {\bibinfo  {journal} {Physical Review B}\ }\textbf {\bibinfo
  {volume} {48}},\ \bibinfo {pages} {11789} (\bibinfo {year}
  {1993})}\BibitemShut {NoStop}%
\bibitem [{\citenamefont {Wondratschek}\ and\ \citenamefont
  {M{\"u}ller}(2004)}]{wondratsheck.2004}%
  \BibitemOpen
  \bibfield  {author} {\bibinfo {author} {\bibfnamefont {H.}~\bibnamefont
  {Wondratschek}}\ and\ \bibinfo {author} {\bibfnamefont {U.}~\bibnamefont
  {M{\"u}ller}},\ }\href@noop {} {\emph {\bibinfo {title} {{International
  Tables for Crystallography: Symmetry Relations between Space Groups}}}},\
  \bibinfo {edition} {1st}\ ed.\ (\bibinfo  {publisher} {Kluwer.},\ \bibinfo
  {year} {2004})\BibitemShut {NoStop}%
\bibitem [{\citenamefont {Monkhorst}\ and\ \citenamefont
  {Pack}(1976)}]{monkhorst.pack.1976}%
  \BibitemOpen
  \bibfield  {author} {\bibinfo {author} {\bibfnamefont {H.~J.}\ \bibnamefont
  {Monkhorst}}\ and\ \bibinfo {author} {\bibfnamefont {J.~D.}\ \bibnamefont
  {Pack}},\ }\href {https://doi.org/10.1103/PhysRevB.13.5188} {\bibfield
  {journal} {\bibinfo  {journal} {Physical Review B}\ }\textbf {\bibinfo
  {volume} {13}},\ \bibinfo {pages} {5188} (\bibinfo {year}
  {1976})}\BibitemShut {NoStop}%
\bibitem [{\citenamefont {Starace}(1971)}]{starace.1971}%
  \BibitemOpen
  \bibfield  {author} {\bibinfo {author} {\bibfnamefont {A.~F.}\ \bibnamefont
  {Starace}},\ }\href {https://doi.org/10.1103/PhysRevA.3.1242} {\bibfield
  {journal} {\bibinfo  {journal} {Physical Review A}\ }\textbf {\bibinfo
  {volume} {3}},\ \bibinfo {pages} {1242} (\bibinfo {year} {1971})}\BibitemShut
  {NoStop}%
\bibitem [{\citenamefont {Lucarini}\ \emph {et~al.}(2005)\citenamefont
  {Lucarini}, \citenamefont {Saarinen}, \citenamefont {Peiponen},\ and\
  \citenamefont {Vartiainen}}]{book.lucarini.2005}%
  \BibitemOpen
  \bibfield  {author} {\bibinfo {author} {\bibfnamefont {V.}~\bibnamefont
  {Lucarini}}, \bibinfo {author} {\bibfnamefont {J.~J.}\ \bibnamefont
  {Saarinen}}, \bibinfo {author} {\bibfnamefont {K.-E.}\ \bibnamefont
  {Peiponen}}, \ and\ \bibinfo {author} {\bibfnamefont {E.~M.}\ \bibnamefont
  {Vartiainen}},\ }\href@noop {} {\emph {\bibinfo {title} {{Kramers-Kronig
  relations in optical materials research}}}},\ Vol.\ \bibinfo {volume} {110}\
  (\bibinfo  {publisher} {Springer Science \& Business Media},\ \bibinfo {year}
  {2005})\BibitemShut {NoStop}%
\bibitem [{\citenamefont {Furuseth}\ \emph {et~al.}(1965)\citenamefont
  {Furuseth}, \citenamefont {Selte}, \citenamefont {Kjekshus}, \citenamefont
  {Gronowitz}, \citenamefont {Hoffman},\ and\ \citenamefont
  {Westerdahl}}]{furuseth.selte.1965}%
  \BibitemOpen
  \bibfield  {author} {\bibinfo {author} {\bibfnamefont {S.}~\bibnamefont
  {Furuseth}}, \bibinfo {author} {\bibfnamefont {K.}~\bibnamefont {Selte}},
  \bibinfo {author} {\bibfnamefont {A.}~\bibnamefont {Kjekshus}}, \bibinfo
  {author} {\bibfnamefont {S.}~\bibnamefont {Gronowitz}}, \bibinfo {author}
  {\bibfnamefont {R.}~\bibnamefont {Hoffman}}, \ and\ \bibinfo {author}
  {\bibfnamefont {A.}~\bibnamefont {Westerdahl}},\ }\href
  {https://doi.org/10.3891/acta.chem.scand.19-0095} {\bibfield  {journal}
  {\bibinfo  {journal} {Acta Chem. Scand.}\ }\textbf {\bibinfo {volume} {19}},\
  \bibinfo {pages} {95} (\bibinfo {year} {1965})}\BibitemShut {NoStop}%
\bibitem [{\citenamefont {Boller}\ and\ \citenamefont
  {Parth{\'e}}(1963)}]{boller.parthe.1963}%
  \BibitemOpen
  \bibfield  {author} {\bibinfo {author} {\bibfnamefont {H.}~\bibnamefont
  {Boller}}\ and\ \bibinfo {author} {\bibfnamefont {E.}~\bibnamefont
  {Parth{\'e}}},\ }\href {https://doi.org/10.1107/S0365110X63002930} {\bibfield
   {journal} {\bibinfo  {journal} {Acta Crystallographica}\ }\textbf {\bibinfo
  {volume} {16}},\ \bibinfo {pages} {1095} (\bibinfo {year}
  {1963})}\BibitemShut {NoStop}%
\bibitem [{\citenamefont {Sch{\"o}nberg}\ \emph {et~al.}(1954)\citenamefont
  {Sch{\"o}nberg}, \citenamefont {Overend}, \citenamefont {Munthe-Kaas},\ and\
  \citenamefont {S{\"o}rensen}}]{schonberg.1954}%
  \BibitemOpen
  \bibfield  {author} {\bibinfo {author} {\bibfnamefont {N.}~\bibnamefont
  {Sch{\"o}nberg}}, \bibinfo {author} {\bibfnamefont {W.}~\bibnamefont
  {Overend}}, \bibinfo {author} {\bibfnamefont {A.}~\bibnamefont
  {Munthe-Kaas}}, \ and\ \bibinfo {author} {\bibfnamefont {N.}~\bibnamefont
  {S{\"o}rensen}},\ }\href {https://doi.org/10.3891/acta.chem.scand.08-0226}
  {\bibfield  {journal} {\bibinfo  {journal} {Acta Chem. Scand}\ }\textbf
  {\bibinfo {volume} {8}} (\bibinfo {year} {1954})}\BibitemShut {NoStop}%
\end{thebibliography}%

\end{document}